\pdfoutput=1
\documentclass[11pt,twoside,a4paper,cmspaper,final,collab]{cms-tdr}
\def\svnVersion{13:19657M}\def\cmsCernNo{2010-039}\def\cmsCernDate{\today}\def\cmsMessage{Submitted to Physics Letters B}

\begin{document}\cmsNoteHeader{TOP-10-001}
%
%
%

%
%
\hyphenation{env-iron-men-tal}
\hyphenation{had-ron-i-za-tion}
\hyphenation{cal-or-i-me-ter}
\hyphenation{de-vices}
%
\RCS$Revision: 19373 $
\RCS$HeadURL: svn+ssh://alverson@svn.cern.ch/reps/tdr2/papers/TOP-10-001/trunk/TOP-10-001.tex $
\RCS$Id: TOP-10-001.tex 19373 2010-10-26 17:12:55Z neu $
%
%
%

\providecommand {\etal}{\mbox{et al.}\xspace} 
\providecommand {\ie}{\mbox{i.e.}\xspace}     
\providecommand {\eg}{\mbox{e.g.}\xspace}     
\providecommand {\etc}{\mbox{etc.}\xspace}     
\providecommand {\vs}{\mbox{\sl vs.}\xspace}      
\providecommand {\mdash}{\ensuremath{\mathrm{-}}} 

\providecommand {\Lone}{Level-1\xspace} 
\providecommand {\Ltwo}{Level-2\xspace}
\providecommand {\Lthree}{Level-3\xspace}

\providecommand{\ACERMC} {\textsc{AcerMC}\xspace}
\providecommand{\ALPGEN} {{\textsc{alpgen}}\xspace}
\providecommand{\CHARYBDIS} {{\textsc{charybdis}}\xspace}
\providecommand{\CMKIN} {\textsc{cmkin}\xspace}
\providecommand{\CMSIM} {{\textsc{cmsim}}\xspace}
\providecommand{\CMSSW} {{\textsc{cmssw}}\xspace}
\providecommand{\COBRA} {{\textsc{cobra}}\xspace}
\providecommand{\COCOA} {{\textsc{cocoa}}\xspace}
\providecommand{\COMPHEP} {\textsc{CompHEP}\xspace}
\providecommand{\EVTGEN} {{\textsc{evtgen}}\xspace}
\providecommand{\FAMOS} {{\textsc{famos}}\xspace}
\providecommand{\GARCON} {\textsc{garcon}\xspace}
\providecommand{\GARFIELD} {{\textsc{garfield}}\xspace}
\providecommand{\GEANE} {{\textsc{geane}}\xspace}
\providecommand{\GEANTfour} {{\textsc{geant4}}\xspace}
\providecommand{\GEANTthree} {{\textsc{geant3}}\xspace}
\providecommand{\GEANT} {{\textsc{geant}}\xspace}
\providecommand{\HDECAY} {\textsc{hdecay}\xspace}
\providecommand{\HERWIG} {{\textsc{herwig}}\xspace}
\providecommand{\HIGLU} {{\textsc{higlu}}\xspace}
\providecommand{\HIJING} {{\textsc{hijing}}\xspace}
\providecommand{\IGUANA} {\textsc{iguana}\xspace}
\providecommand{\ISAJET} {{\textsc{isajet}}\xspace}
\providecommand{\ISAPYTHIA} {{\textsc{isapythia}}\xspace}
\providecommand{\ISASUGRA} {{\textsc{isasugra}}\xspace}
\providecommand{\ISASUSY} {{\textsc{isasusy}}\xspace}
\providecommand{\ISAWIG} {{\textsc{isawig}}\xspace}
\providecommand{\MADGRAPH} {\textsc{MadGraph}\xspace}
\providecommand{\MCATNLO} {\textsc{mc@nlo}\xspace}
\providecommand{\MCFM} {\textsc{mcfm}\xspace}
\providecommand{\MILLEPEDE} {{\textsc{millepede}}\xspace}
\providecommand{\ORCA} {{\textsc{orca}}\xspace}
\providecommand{\OSCAR} {{\textsc{oscar}}\xspace}
\providecommand{\PHOTOS} {\textsc{photos}\xspace}
\providecommand{\PROSPINO} {\textsc{prospino}\xspace}
\providecommand{\PYTHIA} {{\textsc{pythia}}\xspace}
\providecommand{\SHERPA} {{\textsc{sherpa}}\xspace}
\providecommand{\TAUOLA} {\textsc{tauola}\xspace}
\providecommand{\TOPREX} {\textsc{TopReX}\xspace}
\providecommand{\XDAQ} {{\textsc{xdaq}}\xspace}

\providecommand {\DZERO}{D\O\xspace}     


\providecommand{\de}{\ensuremath{^\circ}}
\providecommand{\ten}[1]{\ensuremath{\times \text{10}^\text{#1}}}
\providecommand{\unit}[1]{\ensuremath{\text{\,#1}}\xspace}
\providecommand{\mum}{\ensuremath{\,\mu\text{m}}\xspace}
\providecommand{\micron}{\ensuremath{\,\mu\text{m}}\xspace}
\providecommand{\cm}{\ensuremath{\,\text{cm}}\xspace}
\providecommand{\mm}{\ensuremath{\,\text{mm}}\xspace}
\providecommand{\mus}{\ensuremath{\,\mu\text{s}}\xspace}
\providecommand{\keV}{\ensuremath{\,\text{ke\hspace{-.08em}V}}\xspace}
\providecommand{\MeV}{\ensuremath{\,\text{Me\hspace{-.08em}V}}\xspace}
\providecommand{\GeV}{\ensuremath{\,\text{Ge\hspace{-.08em}V}}\xspace}
\providecommand{\TeV}{\ensuremath{\,\text{Te\hspace{-.08em}V}}\xspace}
\providecommand{\PeV}{\ensuremath{\,\text{Pe\hspace{-.08em}V}}\xspace}
\providecommand{\keVc}{\ensuremath{{\,\text{ke\hspace{-.08em}V\hspace{-0.16em}/\hspace{-0.08em}}c}}\xspace}
\providecommand{\MeVc}{\ensuremath{{\,\text{Me\hspace{-.08em}V\hspace{-0.16em}/\hspace{-0.08em}}c}}\xspace}
\providecommand{\GeVc}{\ensuremath{{\,\text{Ge\hspace{-.08em}V\hspace{-0.16em}/\hspace{-0.08em}}c}}\xspace}
\providecommand{\TeVc}{\ensuremath{{\,\text{Te\hspace{-.08em}V\hspace{-0.16em}/\hspace{-0.08em}}c}}\xspace}
\providecommand{\keVcc}{\ensuremath{{\,\text{ke\hspace{-.08em}V\hspace{-0.16em}/\hspace{-0.08em}}c^\text{2}}}\xspace}
\providecommand{\MeVcc}{\ensuremath{{\,\text{Me\hspace{-.08em}V\hspace{-0.16em}/\hspace{-0.08em}}c^\text{2}}}\xspace}
\providecommand{\GeVcc}{\ensuremath{{\,\text{Ge\hspace{-.08em}V\hspace{-0.16em}/\hspace{-0.08em}}c^\text{2}}}\xspace}
\providecommand{\TeVcc}{\ensuremath{{\,\text{Te\hspace{-.08em}V\hspace{-0.16em}/\hspace{-0.08em}}c^\text{2}}}\xspace}

\providecommand{\pbinv} {\mbox{\ensuremath{\,\text{pb}^\text{$-$1}}}\xspace}
\providecommand{\fbinv} {\mbox{\ensuremath{\,\text{fb}^\text{$-$1}}}\xspace}
\providecommand{\nbinv} {\mbox{\ensuremath{\,\text{nb}^\text{$-$1}}}\xspace}
\providecommand{\percms}{\ensuremath{\,\text{cm}^\text{$-$2}\,\text{s}^\text{$-$1}}\xspace}
\providecommand{\lumi}{\ensuremath{\mathcal{L}}\xspace}
\providecommand{\Lumi}{\ensuremath{\mathcal{L}}\xspace}
%
\providecommand{\LvLow}  {\ensuremath{\mathcal{L}=\text{10}^\text{32}\,\text{cm}^\text{$-$2}\,\text{s}^\text{$-$1}}\xspace}
\providecommand{\LLow}   {\ensuremath{\mathcal{L}=\text{10}^\text{33}\,\text{cm}^\text{$-$2}\,\text{s}^\text{$-$1}}\xspace}
\providecommand{\lowlumi}{\ensuremath{\mathcal{L}=\text{2}\times \text{10}^\text{33}\,\text{cm}^\text{$-$2}\,\text{s}^\text{$-$1}}\xspace}
\providecommand{\LMed}   {\ensuremath{\mathcal{L}=\text{2}\times \text{10}^\text{33}\,\text{cm}^\text{$-$2}\,\text{s}^\text{$-$1}}\xspace}
\providecommand{\LHigh}  {\ensuremath{\mathcal{L}=\text{10}^\text{34}\,\text{cm}^\text{$-$2}\,\text{s}^\text{$-$1}}\xspace}
\providecommand{\hilumi} {\ensuremath{\mathcal{L}=\text{10}^\text{34}\,\text{cm}^\text{$-$2}\,\text{s}^\text{$-$1}}\xspace}


\providecommand{\zp}{\ensuremath{\mathrm{Z}^\prime}\xspace}


\providecommand{\kt}{\ensuremath{k_{\mathrm{T}}}\xspace}
\providecommand{\BC}{\ensuremath{\mathrm{B_{c}}}\xspace}
\providecommand{\bbarc}{\ensuremath{\mathrm{\overline{b}c}}\xspace}
\providecommand{\bbbar}{\ensuremath{\mathrm{b\overline{b}}}\xspace}
\providecommand{\ccbar}{\ensuremath{\mathrm{c\overline{c}}}\xspace}
\providecommand{\JPsi}{\ensuremath{\mathrm{J}\hspace{-.08em}/\hspace{-.14em}\psi}\xspace}
\providecommand{\bspsiphi}{\ensuremath{\mathrm{B_s} \to \JPsi\, \phi}\xspace}
\providecommand{\AFB}{\ensuremath{A_\text{FB}}\xspace}
\providecommand{\EE}{\ensuremath{\mathrm{e^+e^-}}\xspace}
\providecommand{\MM}{\ensuremath{\mu^+\mu^-}\xspace}
\providecommand{\TT}{\ensuremath{\tau^+\tau^-}\xspace}
\providecommand{\wangle}{\ensuremath{\sin^{2}\theta_{\text{eff}}^\text{lept}(M^2_\mathrm{Z})}\xspace}
\providecommand{\ttbar}{\ensuremath{\mathrm{t\overline{t}}}\xspace}
\providecommand{\stat}{\ensuremath{\,\text{(stat.)}}\xspace}
\providecommand{\syst}{\ensuremath{\,\text{(syst.)}}\xspace}

\providecommand{\HGG}{\ensuremath{\mathrm{H}\to\gamma\gamma}}
\providecommand{\gev}{\GeV}
\providecommand{\GAMJET}{\ensuremath{\gamma + \text{jet}}}
\providecommand{\PPTOJETS}{\ensuremath{\mathrm{pp}\to\text{jets}}}
\providecommand{\PPTOGG}{\ensuremath{\mathrm{pp}\to\gamma\gamma}}
\providecommand{\PPTOGAMJET}{\ensuremath{\mathrm{pp}\to\gamma + \mathrm{jet}}}
\providecommand{\MH}{\ensuremath{M_{\mathrm{H}}}}
\providecommand{\RNINE}{\ensuremath{R_\mathrm{9}}}
\providecommand{\DR}{\ensuremath{\Delta R}}


\providecommand{\PT}{\ensuremath{p_{\mathrm{T}}}\xspace}
\providecommand{\pt}{\ensuremath{p_{\mathrm{T}}}\xspace}
\providecommand{\ET}{\ensuremath{E_{\mathrm{T}}}\xspace}
\providecommand{\HT}{\ensuremath{H_{\mathrm{T}}}\xspace}
\providecommand{\et}{\ensuremath{E_{\mathrm{T}}}\xspace}
\providecommand{\Em}{\ensuremath{E\hspace{-0.6em}/}\xspace}
\providecommand{\Pm}{\ensuremath{p\hspace{-0.5em}/}\xspace}
\providecommand{\PTm}{\ensuremath{{p}_\mathrm{T}\hspace{-1.02em}/}\xspace}
\providecommand{\PTslash}{\ensuremath{{p}_\mathrm{T}\hspace{-1.02em}/}\xspace}
\providecommand{\ETm}{\ensuremath{E_{\mathrm{T}}^{\text{miss}}}\xspace}
\providecommand{\ETslash}{\ensuremath{E_{\mathrm{T}}\hspace{-1.1em}/}\xspace}
\providecommand{\MET}{\ensuremath{E_{\mathrm{T}}^{\text{miss}}}\xspace}
\providecommand{\ETmiss}{\ensuremath{E_{\mathrm{T}}^{\text{miss}}}\xspace}
\providecommand{\VEtmiss}{\ensuremath{{\vec E}_{\mathrm{T}}^{\text{miss}}}\xspace}

\providecommand{\dd}[2]{\ensuremath{\frac{\mathrm{d} #1}{\mathrm{d} #2}}}

%

\providecommand{\ga}{\ensuremath{\gtrsim}}
\providecommand{\la}{\ensuremath{\lesssim}}
\providecommand{\swsq}{\ensuremath{\sin^2\theta_\mathrm{W}}\xspace}
\providecommand{\cwsq}{\ensuremath{\cos^2\theta_\mathrm{W}}\xspace}
\providecommand{\tanb}{\ensuremath{\tan\beta}\xspace}
\providecommand{\tanbsq}{\ensuremath{\tan^{2}\beta}\xspace}
\providecommand{\sidb}{\ensuremath{\sin 2\beta}\xspace}
\providecommand{\alpS}{\ensuremath{\alpha_S}\xspace}
\providecommand{\alpt}{\ensuremath{\tilde{\alpha}}\xspace}

\providecommand{\QL}{\ensuremath{\mathrm{Q}_\mathrm{L}}\xspace}
\providecommand{\sQ}{\ensuremath{\tilde{\mathrm{Q}}}\xspace}
\providecommand{\sQL}{\ensuremath{\tilde{\mathrm{Q}}_\mathrm{L}}\xspace}
\providecommand{\ULC}{\ensuremath{\mathrm{U}_\mathrm{L}^\mathrm{C}}\xspace}
\providecommand{\sUC}{\ensuremath{\tilde{\mathrm{U}}^\mathrm{C}}\xspace}
\providecommand{\sULC}{\ensuremath{\tilde{\mathrm{U}}_\mathrm{L}^\mathrm{C}}\xspace}
\providecommand{\DLC}{\ensuremath{\mathrm{D}_\mathrm{L}^\mathrm{C}}\xspace}
\providecommand{\sDC}{\ensuremath{\tilde{\mathrm{D}}^\mathrm{C}}\xspace}
\providecommand{\sDLC}{\ensuremath{\tilde{\mathrm{D}}_\mathrm{L}^\mathrm{C}}\xspace}
\providecommand{\LL}{\ensuremath{\mathrm{L}_\mathrm{L}}\xspace}
\providecommand{\sL}{\ensuremath{\tilde{\mathrm{L}}}\xspace}
\providecommand{\sLL}{\ensuremath{\tilde{\mathrm{L}}_\mathrm{L}}\xspace}
\providecommand{\ELC}{\ensuremath{\mathrm{E}_\mathrm{L}^\mathrm{C}}\xspace}
\providecommand{\sEC}{\ensuremath{\tilde{\mathrm{E}}^\mathrm{C}}\xspace}
\providecommand{\sELC}{\ensuremath{\tilde{\mathrm{E}}_\mathrm{L}^\mathrm{C}}\xspace}
\providecommand{\sEL}{\ensuremath{\tilde{\mathrm{E}}_\mathrm{L}}\xspace}
\providecommand{\sER}{\ensuremath{\tilde{\mathrm{E}}_\mathrm{R}}\xspace}
\providecommand{\sFer}{\ensuremath{\tilde{\mathrm{f}}}\xspace}
\providecommand{\sQua}{\ensuremath{\tilde{\mathrm{q}}}\xspace}
\providecommand{\sUp}{\ensuremath{\tilde{\mathrm{u}}}\xspace}
\providecommand{\suL}{\ensuremath{\tilde{\mathrm{u}}_\mathrm{L}}\xspace}
\providecommand{\suR}{\ensuremath{\tilde{\mathrm{u}}_\mathrm{R}}\xspace}
\providecommand{\sDw}{\ensuremath{\tilde{\mathrm{d}}}\xspace}
\providecommand{\sdL}{\ensuremath{\tilde{\mathrm{d}}_\mathrm{L}}\xspace}
\providecommand{\sdR}{\ensuremath{\tilde{\mathrm{d}}_\mathrm{R}}\xspace}
\providecommand{\sTop}{\ensuremath{\tilde{\mathrm{t}}}\xspace}
\providecommand{\stL}{\ensuremath{\tilde{\mathrm{t}}_\mathrm{L}}\xspace}
\providecommand{\stR}{\ensuremath{\tilde{\mathrm{t}}_\mathrm{R}}\xspace}
\providecommand{\stone}{\ensuremath{\tilde{\mathrm{t}}_1}\xspace}
\providecommand{\sttwo}{\ensuremath{\tilde{\mathrm{t}}_2}\xspace}
\providecommand{\sBot}{\ensuremath{\tilde{\mathrm{b}}}\xspace}
\providecommand{\sbL}{\ensuremath{\tilde{\mathrm{b}}_\mathrm{L}}\xspace}
\providecommand{\sbR}{\ensuremath{\tilde{\mathrm{b}}_\mathrm{R}}\xspace}
\providecommand{\sbone}{\ensuremath{\tilde{\mathrm{b}}_1}\xspace}
\providecommand{\sbtwo}{\ensuremath{\tilde{\mathrm{b}}_2}\xspace}
\providecommand{\sLep}{\ensuremath{\tilde{\mathrm{l}}}\xspace}
\providecommand{\sLepC}{\ensuremath{\tilde{\mathrm{l}}^\mathrm{C}}\xspace}
\providecommand{\sEl}{\ensuremath{\tilde{\mathrm{e}}}\xspace}
\providecommand{\sElC}{\ensuremath{\tilde{\mathrm{e}}^\mathrm{C}}\xspace}
\providecommand{\seL}{\ensuremath{\tilde{\mathrm{e}}_\mathrm{L}}\xspace}
\providecommand{\seR}{\ensuremath{\tilde{\mathrm{e}}_\mathrm{R}}\xspace}
\providecommand{\snL}{\ensuremath{\tilde{\nu}_L}\xspace}
\providecommand{\sMu}{\ensuremath{\tilde{\mu}}\xspace}
\providecommand{\sNu}{\ensuremath{\tilde{\nu}}\xspace}
\providecommand{\sTau}{\ensuremath{\tilde{\tau}}\xspace}
\providecommand{\Glu}{\ensuremath{\mathrm{g}}\xspace}
\providecommand{\sGlu}{\ensuremath{\tilde{\mathrm{g}}}\xspace}
\providecommand{\Wpm}{\ensuremath{\mathrm{W}^{\pm}}\xspace}
\providecommand{\sWpm}{\ensuremath{\tilde{\mathrm{W}}^{\pm}}\xspace}
\providecommand{\Wz}{\ensuremath{\mathrm{W}^{0}}\xspace}
\providecommand{\sWz}{\ensuremath{\tilde{\mathrm{W}}^{0}}\xspace}
\providecommand{\sWino}{\ensuremath{\tilde{\mathrm{W}}}\xspace}
\providecommand{\Bz}{\ensuremath{\mathrm{B}^{0}}\xspace}
\providecommand{\sBz}{\ensuremath{\tilde{\mathrm{B}}^{0}}\xspace}
\providecommand{\sBino}{\ensuremath{\tilde{\mathrm{B}}}\xspace}
\providecommand{\Zz}{\ensuremath{\mathrm{Z}^{0}}\xspace}
\providecommand{\sZino}{\ensuremath{\tilde{\mathrm{Z}}^{0}}\xspace}
\providecommand{\sGam}{\ensuremath{\tilde{\gamma}}\xspace}
\providecommand{\chiz}{\ensuremath{\tilde{\chi}^{0}}\xspace}
\providecommand{\chip}{\ensuremath{\tilde{\chi}^{+}}\xspace}
\providecommand{\chim}{\ensuremath{\tilde{\chi}^{-}}\xspace}
\providecommand{\chipm}{\ensuremath{\tilde{\chi}^{\pm}}\xspace}
\providecommand{\Hone}{\ensuremath{\mathrm{H}_\mathrm{d}}\xspace}
\providecommand{\sHone}{\ensuremath{\tilde{\mathrm{H}}_\mathrm{d}}\xspace}
\providecommand{\Htwo}{\ensuremath{\mathrm{H}_\mathrm{u}}\xspace}
\providecommand{\sHtwo}{\ensuremath{\tilde{\mathrm{H}}_\mathrm{u}}\xspace}
\providecommand{\sHig}{\ensuremath{\tilde{\mathrm{H}}}\xspace}
\providecommand{\sHa}{\ensuremath{\tilde{\mathrm{H}}_\mathrm{a}}\xspace}
\providecommand{\sHb}{\ensuremath{\tilde{\mathrm{H}}_\mathrm{b}}\xspace}
\providecommand{\sHpm}{\ensuremath{\tilde{\mathrm{H}}^{\pm}}\xspace}
\providecommand{\hz}{\ensuremath{\mathrm{h}^{0}}\xspace}
\providecommand{\Hz}{\ensuremath{\mathrm{H}^{0}}\xspace}
\providecommand{\Az}{\ensuremath{\mathrm{A}^{0}}\xspace}
\providecommand{\Hpm}{\ensuremath{\mathrm{H}^{\pm}}\xspace}
\providecommand{\sGra}{\ensuremath{\tilde{\mathrm{G}}}\xspace}
\providecommand{\mtil}{\ensuremath{\tilde{m}}\xspace}
\providecommand{\rpv}{\ensuremath{\rlap{\kern.2em/}R}\xspace}
\providecommand{\LLE}{\ensuremath{LL\bar{E}}\xspace}
\providecommand{\LQD}{\ensuremath{LQ\bar{D}}\xspace}
\providecommand{\UDD}{\ensuremath{\overline{UDD}}\xspace}
\providecommand{\Lam}{\ensuremath{\lambda}\xspace}
\providecommand{\Lamp}{\ensuremath{\lambda'}\xspace}
\providecommand{\Lampp}{\ensuremath{\lambda''}\xspace}
\providecommand{\spinbd}[2]{\ensuremath{\bar{#1}_{\dot{#2}}}\xspace}

\providecommand{\MD}{\ensuremath{{M_\mathrm{D}}}\xspace}
\providecommand{\Mpl}{\ensuremath{{M_\mathrm{Pl}}}\xspace}
\providecommand{\Rinv} {\ensuremath{{R}^{-1}}\xspace} 
\def\mrm{\mathrm}
\def\ra{\rightarrow}

\newcommand{\roots}{\ensuremath{\sqrt{s}}}
\newcommand{\lhcE}[1]{\ensuremath{\roots ={#1}~\TeV}}

\renewcommand{\sign}{\ensuremath{\mathrm{sign}}}
\newcommand{\rphi}{\text{$r$-$\phi$}}
\newcommand{\etaphi}{\text{$\eta$-$\phi$}}
\newcommand{\rz}{\text{$r$-$z$}}
\newcommand{\met} {\ensuremath{E\!\!\!\!/_T}}
\newcommand{\jpt}{\ensuremath{\mathrm{JPT}}}
\newcommand{\tcmet}{\ensuremath{\mrm{tcMET}}}
\newcommand{\calomet}{\ensuremath{\mrm{\met^{\mrm{calo,Type1}}}}}
\newcommand{\pfmet}{\ensuremath{\mrm{\met^{\mrm{PF}}}}}
\newcommand{\jet}{\ensuremath{\mrm{jet}}}
\newcommand{\jets}{\ensuremath{\mrm{jets}}}
\newcommand{\njet}{\ensuremath{N_\jet}}
\newcommand{\isotk}{\ensuremath{I_\mrm{trk}}}
\newcommand{\isocal}{\ensuremath{I_\mrm{cal}}}
\newcommand{\isocomb}{\ensuremath{I_\mrm{comb}}}
\newcommand{\mll}{\ensuremath{M_{\ell\ell}}}
\newcommand{\etsc}{\ensuremath{E_{T}^{\mrm{sc}}}}
\newcommand{\dxybs}{\ensuremath{d_0^{\mrm{BS}}}}

\newcommand{\jp}{\ensuremath{J/\psi}}
\newcommand{\eepm}{\ensuremath{e^+ e^-}}
\newcommand{\mmpm}{\ensuremath{\mu^+ \mu^-}}
\newcommand{\ttpm}{\ensuremath{\tau^+ \tau^-}}
\newcommand{\empm}{\ensuremath{e^\pm \mu^\mp}}
\newcommand{\dy}{\ensuremath{Z/\gamma^\star}}
\newcommand{\dyee}{\ensuremath{\dy\to\eepm}}
\newcommand{\dymm}{\ensuremath{\dy\to\mmpm}}
\newcommand{\dytt}{\ensuremath{\dy\to\ttpm}}
\newcommand{\Zee}{\ensuremath{Z\to\eepm}}
\newcommand{\Zmm}{\ensuremath{Z\to\mmpm}}
\newcommand{\wen}{\ensuremath{W\to e\nu_{e}}}
\newcommand{\wmn}{\ensuremath{W\to\mu\nu_{\mu}}}
\newcommand{\wtn}{\ensuremath{W\to\tau\nu_{\tau}}}

\newcommand{\EEE}[1]{\ensuremath{\times 10^{#1}}}
\newcommand{\pp}{\ensuremath{pp}}

\newcommand{\totLumi}{\ensuremath{2.79~\pbinv}}

\newcommand{\nNoNo}{\ensuremath{N_{\overline{n}\overline{n}}}}
\newcommand{\nNoNu}{\ensuremath{N_{{n}\overline{n}}}}
\newcommand{\nNuNu}{\ensuremath{N_{{n}{n}}}}

\newcommand{\sye}[1]{\ensuremath{~\pm #1}}
\newcommand{\ase}[2]{\ensuremath{_{~- #1}^{~+ #2}}}
\newcommand{\asi}[2]{\ensuremath{_{~- #1}^{~+ #2}}}

\cmsNoteHeader{TOP-10-001} 
\title{First Measurement of the Cross Section for Top-Quark Pair Production in Proton-Proton Collisions at $\sqrt{s}=7\,\rm{TeV}$}

\address[cms]{CMS}
\author[cms]{The CMS Collaboration}

\date{\today}


\abstract{
The first measurement of the cross section for top-quark pair
production in pp collisions at the LHC at center-of-mass energy
$\sqrt{s}= 7 \rm\ TeV$ has been performed using ${\rm 3.1 \pm 0.3}$~${\rm
pb^{-1}}$ of data recorded by the CMS detector.  This result utilizes
the final state with two isolated, highly energetic charged leptons,
large missing transverse energy, and two or more jets.  Backgrounds
from Drell-Yan and non-W/Z boson production are estimated from
data. Eleven events are observed in the data with ${\rm 2.1 \pm 1.0}$
events expected from background.  The measured cross section is ${\rm
194 \pm 72 {\rm (stat.)} \pm 24 {\rm (syst.)} \pm 21 {\rm
(lumi.)}}$~pb, consistent with next-to-leading order predictions.  }

\hypersetup{%
pdfauthor={CMS Collaboration},%
pdftitle={First Measurement of the Cross Section for Top-Quark Pair Production in Proton-Proton Collisions at 7 TeV},%
pdfsubject={CMS},%
pdfkeywords={CMS, LHC, physics, top quark, cross section}}

\maketitle 


\newcommand{\EM}{\ensuremath{e^{\pm}\mu^{\mp}}\xspace}
\newcommand{\mymet}{\makebox[2.4ex]{\ensuremath{\not\!\! E_{\mathrm{T}}}}}
\newcommand{\myttbar}{\ensuremath{{t\overline{t}}}\xspace}

Since its discovery~\cite{top-discovery2,top-discovery3}, the
properties of the top quark have been subject to numerous detailed
studies~\cite{Incandela:2009pf}, which until recently have only been
possible at the Tevatron proton-antiproton collider.  With the advent
of the Large Hadron Collider (LHC) era~\cite{lhc}, top-quark processes
can be studied for the first time in multi-TeV proton-proton
collisions.  In both p${\rm \overline{p}}$ and pp collisions, top
quarks are expected to be produced primarily via the strong
interaction in top-antitop (${\rm t{\bar t}}$) pairs.  At the
LHC, the ${\rm t{\bar t}}$ production mechanism is expected to be
dominated by a gluon fusion process, whereas at the Tevatron, top-quark
pairs are predominantly produced through quark-antiquark annihilation.
Measurements of top-quark production at the LHC are therefore
important new tests of our understanding of the ${\rm t{\bar t}}$
production mechanism.  This is a crucial component of the early LHC
physics program, since many signatures of new physics models
accessible at the LHC either suffer from top-quark production as a
significant background or contain top quarks themselves.

In this Letter we present the first measurement of the cross section
for ${\rm t{\bar t}}$ production in proton-proton collisions at the
LHC at center-of-mass energy $\sqrt{s}$ = 7 TeV.  The results are
based on a data sample corresponding to an integrated luminosity of
${\rm 3.1 \pm 0.3}$~\pbinv \cite{lumipas} recorded by the CMS
experiment~\cite{JINST} between March and August 2010.  This
measurement is an important milestone for CMS,
demonstrating the experiment's capabilities in extracting an intricate
signature.

Within the standard model, the top quark decays via the weak process
${\rm t\rightarrow Wb}$ almost exclusively.  Experimentally, top-quark
pair events are categorized according to the decay of the two ${\rm
W}$ bosons: the all-hadronic channel, in which both ${\rm W}$ bosons
decay into quarks; the lepton+jets channel, in which one ${\rm W}$
boson decays leptonically, the other into quarks; and the dilepton
channel, in which both ${\rm W}$ bosons decay into leptons.  The
measurement described herein is performed using the ${\rm e^+e^-}$,
\Pgmp \Pgmm , and ${\rm e^{\pm}}$\Pgm $^{\mp}$ dilepton ${\rm t{\bar 
t}}$ modes.
Therefore, the final state studied in this analysis contains two
oppositely charged leptons (electrons and muons, including taus
subsequently decaying to electrons and muons), two neutrinos from
the ${\rm W}$ boson decays, and at least two jets of particles
resulting from the hadronization of the ${\rm b}$ quarks.

The central feature of the CMS apparatus is a superconducting solenoid,
13~m in length and 6~m in diameter, which provides an axial magnetic
field of 3.8~T.  The bore of the solenoid is outfitted with various
particle detection systems.  Charged particle trajectories are
measured by the silicon pixel and strip tracker, covering $0 < \phi <
2\pi$ in azimuth and $|\eta |<$~2.5, where the pseudorapidity $\eta$ is
defined as $\eta =-\ln[\tan{\theta/2}]$, with $\theta$ being the
polar angle of the trajectory of the particle with respect to the
beam direction.  A crystal electromagnetic calorimeter
(ECAL) and a brass/scintillator hadronic calorimeter (HCAL) surround
the tracking volume;
in this analysis the calorimetry provides high-resolution energy and
direction measurements of electrons and hadronic jets.  Muons are measured in
gas detectors embedded in the steel return yoke outside the solenoid.
The detector is nearly hermetic, allowing for energy balance
measurements in the plane transverse to the beam directions.  A
two-tier trigger system selects the most interesting pp collision
events for use in physics analysis.  A more detailed description of
the CMS detector can be found elsewhere~\cite{JINST}.

The trigger providing the data sample used in this analysis
is based on the presence of at least one charged lepton, either an
electron or a muon, with a minimum transverse momentum $p_T$ of
9 (15) \GeVc for the muon (electron).  This data sample is
used both for the selection of the signal and for signal-depleted
control regions used for studies related to background
processes.
Simulated signal events that pass the event selection, as described
below, satisfy the trigger requirements with an efficiency above 97\% in
the $\Pgmp\Pgmm$ decay mode and above 99\% in the other two modes, in
agreement with estimates from Z boson events in the data.

Before
further consideration, events are required to have at least one good
reconstructed pp interaction vertex~\cite{trkpas}.
Among these events, selection criteria are
applied to reconstructed objects to identify candidates consistent
with dilepton ${\rm t{\bar t}}$ processes.

Muon candidates are reconstructed~\cite{MUOPAS} using two algorithms
that require consistent hits in the tracker and muon systems: one is
an algorithm based on the matching of extrapolated trajectories from
the silicon tracker to hits in the muon system (tracker-based muons);
the second is an algorithm based on performing a global fit of
consistent hits in the tracker and the muon system (globally-fitted
muons).
Candidates are required 
to have $p_T>$~20~\GeVc\ and $|\eta|<$~2.5.
Additionally, the track associated with the muon candidate is required
to have a minimum number of hits in the silicon tracker, to be
consistent with originating from the beam spot, and to have a
high-quality global fit including a minimum number of hits in the muon
detector.

Electron candidates are reconstructed~\cite{EGMPAS} starting from a
cluster of energy deposits in the crystals of the ECAL, which is then
matched to hits in the silicon tracker, used to initiate a special
track reconstruction algorithm.  The electron reconstruction algorithm
takes into account the possibility of significant energy loss of the
electron through bremsstrahlung as it traverses the material of the
tracker.  Electron candidates are required to have $p_T>$~20~\GeVc\ and pseudorapidity $|\eta|<$~2.5.  The electron
candidate track is required to be consistent with originating from the
beam spot.  Requirements on the values of electron identification
variables based on shower shape and track-cluster matching are applied
to the reconstructed candidates; the criteria are optimized in the
context of the inclusive ${\rm W}\rightarrow {\rm e}\nu$ selection
and are designed to maximize the rejection of electron candidates from
QCD multijet production while maintaining 90\% efficiency for
electrons from the decay of ${\rm W/Z}$ bosons.  Electron candidates
within $\Delta R=\sqrt{\Delta\phi^2+\Delta\eta^2}<$~0.1 of a
tracker-based or globally-fitted muon are rejected to remove fake
electron candidates due to muon bremsstrahlung.  In addition,
electrons consistent with anomalous depositions in the ECAL or with
photon conversions are rejected.

Charged leptons from the decay of ${\rm W}$ bosons are expected to be
isolated from other activity in the event.  For selected muon and
electron candidates, a cone of $\Delta R <$~0.3 is constructed
around the track direction at the origin and the scalar sum of the
track transverse momenta and calorimeter energy deposits projected onto
a plane transverse to the beam is calculated.  The contribution from
the candidate itself is excluded.  If the value of this scalar sum is
more than 15\% of the candidate's transverse momentum, the
candidate is considered to be non-isolated and is rejected.  

Lepton trigger, identification, and isolation efficiencies are
measured using inclusive Z events from data and are compared with
simulation.  All comparisons show good agreement, generally within
2\%. The residual differences between the efficiencies estimated in
data and simulation are treated as systematic uncertainties.

Events are required to have at least one pair of oppositely charged
leptons.  Both charged leptons are required to originate from within
1~cm along the beamline of the reconstructed pp interaction location.
To veto contributions from ${\rm Z}$ production, the
invariant mass of the dilepton system, $\mll$, is required to be
outside a $\pm$15~\GeVcc\ window centered at the mass of the ${\rm Z}$ boson for
the ${\rm e^+e^-}$ and $\Pgmp\Pgmm$ modes.  Additionally, dilepton
candidate events with $\mll <$~10~\GeVcc\ are removed, at
essentially no penalty for the collected signal.

The neutrinos from the ${\rm W}$ boson decays do not interact with the
detector and escape without depositing any of their energy.  The presence
of a neutrino manifests itself as an imbalance in the measured energy
depositions; the imbalance in the projection perpendicular to the beam
line (missing transverse energy, $\met$) is an important distinguishing feature of ${\rm t{\bar t}}$
events in this channel.  At CMS there are several techniques for
calculating $\met$~\cite{METPAS2}; here, the raw $\met$,
calculated from calorimeter signals, is made more accurate through a
series of corrections taking into account the contribution from the
minimally interacting muons and, most importantly, a per-track
correction for the expected imperfect response of the calorimeter.
This track correction results in an improved energy resolution,
especially for low-energy charged particles. Neither the dominant
background processes, Drell-Yan ${\rm Z}/\Pgg^\star \rightarrow {\rm
e^+e^-}$ and $\Pgmp\Pgmm$, nor the
difficult-to-model background from isolated lepton candidates produced
in QCD multijet events, contain a natural source of large $\met$.  Hence, in the ${\rm e^+e^-}$ and $\Pgmp\Pgmm$ modes, $\met$~$>$~30~GeV is required; in the ${\rm e^{\pm}}\Pgm^{\mp}$ mode a
looser requirement of $\met$ $>$~20~GeV is used due to the
significantly smaller contribution of Drell-Yan background.

Dilepton ${\rm t{\bar t}}$ events will have at least two hadronic jets
from the hadronization of the two ${\rm b}$ quarks. The anti-$k_T$ clustering algorithm~\cite{antikt} with $R=0.5$ is used
for jet clustering.  Jets are reconstructed using calorimeter
information and corrected using reconstructed tracks~\cite{JETPAS}.
Further corrections are applied to the raw jet momenta to establish a
relative uniform response of the calorimeter in jet $\eta$ and an
absolute uniform response in jet $p_T$.  The jet energy scale
uncertainty for these track-corrected jets is 5\%.  Jet candidates are
required to have $p_T >$~30~\GeVc , $|\eta|<$~2.5, and must not
overlap with any electron or muon candidate within $\Delta R
<$~0.4.
Events with fewer than two jets are discarded.

The selection efficiency of signal events is evaluated in a simulated
${\rm t{\bar t}}$ event sample modeled with the \MADGRAPH event
generator~\cite{madgraph} with up to three additional hadronic jets.
The events are subsequently processed with \PYTHIA ~\cite{pythia} to
provide showering of generated particles, and then processed with a
full CMS detector simulation based on \GEANTfour ~\cite{geant}.
The total next-to-leading order (NLO) cross section for top-quark pair
production used here to scale simulated signal distributions is
$\sigma_{{\rm \ttbar}}=157.5^{+23.2}_{-24.4} \rm\ pb$, as obtained
with MCFM~\cite{mcfm,mcfm:tt}.  Approximate next-to-next-to-leading
order (NNLO) calculations for the ${\rm t{\bar t}}$ cross section have
been completed (see for example
\cite{Kidonakis:2010bb,Kidonakis:2008mu,Cacciari:2008zb,Moch:2008qy,Langenfeld:2009tc,Langenfeld:2009wd})
but are not used here.  The theoretical uncertainty on the cross
section includes the scale uncertainties, determined by varying the
factorization and renormalization scales by factors of 2 and 0.5
around the central scale, corresponding to the top quark mass
(172.5~\GeVcc ), and the uncertainties from the parton distribution
functions (PDFs) and the value of $\alpha_{\rm S}$, following the
procedures from the MSTW2008~\cite{mstw08}, CTEQ6.6~\cite{cteq66}, and
NNPDF2.0~\cite{nnpdf} sets. 
The expected
yield of events passing the selection criteria,
assuming the NLO production cross section, is ${\rm 1.5 \pm 0.3}$, ${\rm
1.7 \pm 0.3}$, and ${\rm 4.5 \pm 0.9}$ for the ${\rm e^+e^-}$, \Pgmp
\Pgmm , and ${\rm e^{\pm}}$\Pgm $^{\mp}$ decay modes, respectively.
The uncertainties on these predicted event yields combine the
systematic uncertainties on the event selection, the theoretical
production cross section, and the integrated luminosity of the sample,
where the contribution from the last two sources dominates the total.
Note that the simulated ${\rm t{\bar t}}$ signal sample used for the
estimate of the expected signal events was generated with the ${\rm W
\rightarrow \ell}\nu$ branching fraction set to 1/9, which is not 
consistent with the standard value ($0.1080 \pm 0.009$)~\cite{PDG}
used in the cross section measurement.


The selected sample is not 100\% pure in dilepton ${\rm \ttbar}$
events.
There are two types of background estimation techniques used in this
analysis.  One strategy utilizes simulated pp collision events to
model background processes.  There are, however, some pathological
backgrounds that are harder to model accurately.  In such cases, it is
preferred to estimate the yields of these events from the data.

Contributions from diboson production (${\rm VV}$, where ${\rm V = W}$
or ${\rm Z}/\Pgg^\star$), based on a leading-order production cross
section of $\sigma_{{\rm VV}}=4.8$~pb~\cite{madgraph}, and electroweak
single-top production in the ${\rm tW}$ channel ($\sigma_{{\rm
tW}}=10.6$~pb~\cite{mcfm:t:2}) are modeled with the \MADGRAPH\ event
generator and are processed in an equivalent fashion as the simulated
${\rm \ttbar}$ sample used to assess the signal yield.  The Drell-Yan
${\rm Z}/\Pgg^\star \rightarrow \Pgt\Pgt$ process ($\sigma_{{\rm
Z}/\Pgg^\star \rightarrow \Pgt\Pgt}=1666$~pb~\cite{fewz}) is modeled
with \PYTHIA
and \MADGRAPH .  The uncertainties on these production cross sections
are well within the total systematic uncertainty of 50\% used for each
of these backgrounds. Table~\ref{tab:yieldsAllCutsFinal2} gives the
simulation-based predictions for the event yields from these
processes.

\begin{table}[t!]
\begin{center}
\caption{\small\label{tab:yieldsAllCutsFinal2}
The expected number of dilepton ${\rm \ttbar}$ signal and background
events passing the full selection criteria, compared to the number of
observed events.  The procedures for estimating the expected numbers
of events and their uncertainties are described in the text.  The expected 
signal yield assumes a ${\rm \ttbar}$ cross section of 
$\sigma_{{\rm \ttbar}}=157.5^{+23.2}_{-24.4}\rm\ pb$.}
\begin{tabular}{lcc}\\ \hline 
  Source                       & $ $ & Number of events \\ \hline
  Expected ${\rm \ttbar}$      & $ $ & $7.7\pm 1.5$ \\  \hline
  Dibosons (${\rm VV}$)        & $ $ & $0.13\pm 0.07$ \\
  Single top (${\rm tW}$)      & $ $ & $0.25\pm 0.13$ \\
  Drell-Yan ${\rm Z}/\Pgg^\star \rightarrow \Pgt^+\Pgt^-$ & $ $ & $0.18\pm 0.09$ \\
  Drell-Yan ${\rm Z}/\Pgg^\star \rightarrow {\rm e^+e^-}$, $\Pgmp\Pgmm$ & $ $ & $1.4\pm0.5\pm0.5$ \\
  Events with non-W/Z leptons   & $ $ & $0.1\pm0.5\pm0.3$  \\ \hline
  Total backgrounds                                   & $ $ & $2.1\pm1.0$ \\ \hline
  Expected total, including ${\rm \ttbar}$            & $ $ & $9.8\pm1.8$ \\ \hline
  Data                                                & $ $ & 11 \\ \hline
\end{tabular}
\end{center}
\end{table}

The contributions from two important background sources are estimated
from the data: exceptional Drell-Yan events that evade the ${\rm Z}$
veto and are accompanied by significant missing transverse energy; and
non-W/Z isolated lepton signatures from multijet
and ${\rm W}$+jets production.  Difficult-to-simulate instrumental
effects influence both topologies and hence it is preferable to use
calibration samples from the data in these estimations.

The events rejected by the ${\rm Z}$ veto are used to estimate the
residual contributions from Drell-Yan ${\rm Z}/\Pgg^\star \rightarrow
{\rm e^+e^-}$ and $\Pgmp\Pgmm$ in the surviving selected sample.  In
the $\Pgmp\Pgmm$ final state the rate of events surviving the ${\rm
Z}$ veto is equal to an estimate of the Drell-Yan contribution near
the $\mll$ peak, scaled by the expected ratio of off-peak to
near-peak events derived from simulation. The near-peak Drell-Yan
${\rm Z}/\Pgg^\star$ contribution is estimated from the number of all
events triggering the Z veto, after subtraction of the non-Drell-Yan
contribution estimated from ${\rm e^{\pm}}\Pgm^{\mp}$ events passing
the same selection and corrected for the differences between the
electron and muon identification efficiencies.
The estimate in the ${\rm e^{+}e^{-}}$ mode is done in a similar
fashion; the summed contribution is shown in
Table~\ref{tab:yieldsAllCutsFinal2}.  The systematic uncertainty of
this method, evaluated in each mode separately, is estimated to be
50\%.  This is dominated by detector calibration effects and
changes of the fraction of Z-vetoed Drell-Yan ${\rm Z}/\Pgg^\star$
events with increasingly stringent requirements (additional jets
and missing transverse energy) as estimated from simulation.

The contributions to the selected sample from isolated lepton
candidates from non-W/Z decays are also derived from data.  Such
lepton candidates mostly arise from jets that are able to satisfy the
tight lepton identification criteria.  A superset of dilepton
candidate events is chosen by loosening the lepton identification criteria
in the trigger samples used for the measurement.  The number of these
candidates passing the loosened selection criteria from non-W/Z
leptons can be weighted by the ratio of yields of tight-to-loose
lepton candidates ($R_{TL}$) to produce an estimate of non-W/Z
leptons passing the tight selections.  The ratio $R_{TL}$ is
measured as a function of candidate transverse momentum and
pseudorapidity in a multijet-dominated data sample containing 
events with one lepton candidate passing loose selection
criteria.  Additional selection criteria are applied to suppress
the residual contribution to the loose lepton sample from 
electroweak processes.  We assume
this $R_{TL}$ is appropriate for use in the dilepton signal
sample, and we also consider $R_{TL}$ to be independent from the
other lepton in events with two leptons.  Possible differences are
assessed from comparisons of $R_{TL}$ measured in QCD multijet
data samples with different jet thresholds and flavor content.

Estimates for the contributions from lepton candidates in pure multijet
QCD, with two such non-W/Z candidates, and in ${\rm W}$+jets, with one such
candidate beyond that from the decay of the ${\rm W}$, are derived
separately. A sample of loose dilepton events both failing the tight
selections is used to estimate the multijet QCD contribution. Loose
dilepton events with only one lepton failing the tight requirements
include contributions from ${\rm W}$+jets events, but are contaminated by
multijets and leptons from W/Z decays.  The multijet QCD contamination is
subtracted using the previous estimate, while the contamination from
W/Z leptons is measured from a sample of Z events fulfilling loose
selection requirements.

The prediction for these non-W/Z leptons is shown in
Table~\ref{tab:yieldsAllCutsFinal2}.  The systematic uncertainty on
the non-W/Z lepton estimate is primarily from differences in the jet
momentum spectrum and flavor composition between the QCD-dominated
sample in which $R_{TL}$ is measured and the sample where it is
applied.  Other subdominant contributions to the systematic
uncertainties include the $R_{TL}$ measurement biases due to
electroweak signal contribution, the dissimilarity in the trigger
between the $R_{TL}$ calibration sample and the signal sample to
which it is applied, and from the statistical limitations on the
$R_{TL}$ calibration sample.  The systematic uncertainty on the
electron $R_{TL}$ is 50\%, which corresponds to a 50\% (100\%)
uncertainty on a raw estimate of the ${\rm W}$+jets (QCD multijets)
non-W/Z isolated lepton contribution, prior to accounting for the
signal contribution to the estimate.  Similarly, the systematic
uncertainty on the muon $R_{TL}$ is ${\rm \ase{100}{50}\%}$,
which corresponds to a ${\rm \ase{100}{50}\%}$ (${\rm
\ase{100}{100}\%}$) uncertainty on the estimate of the ${\rm W}$+jets
(QCD multijets) non-W/Z isolated lepton contribution.

\begin{figure}[t!]
\begin{center}
  \includegraphics[width=0.45\textwidth]{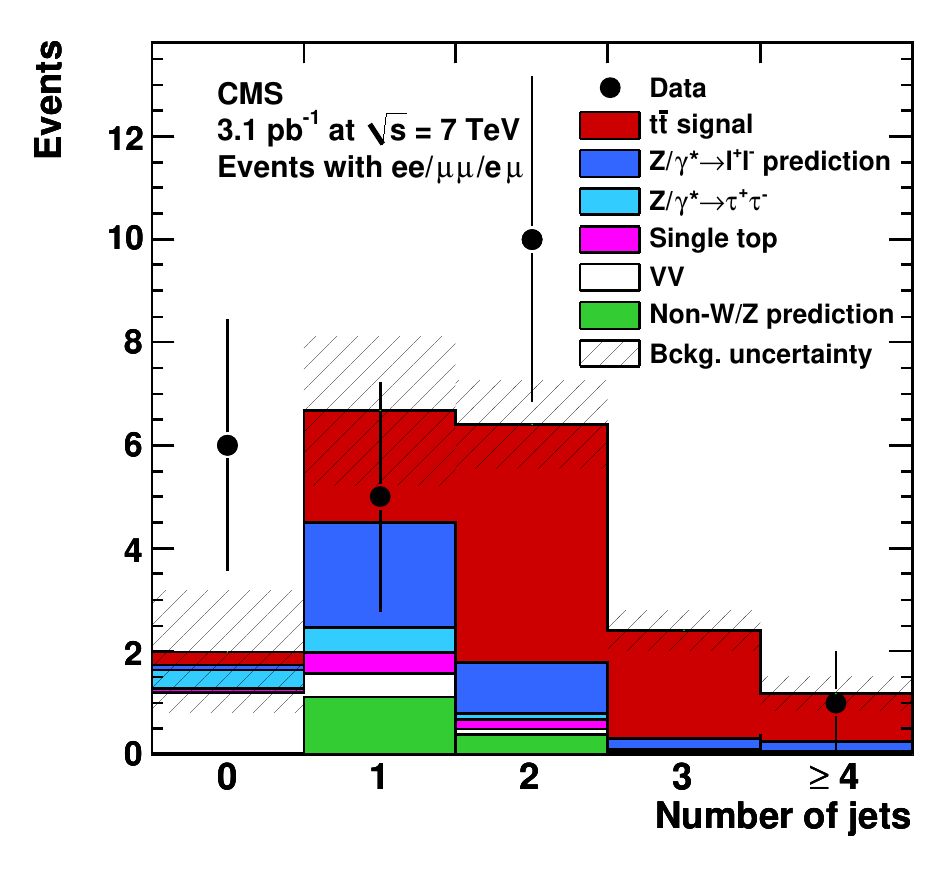}
\caption{\small Number of jets in events passing
all dilepton selection criteria before the $\geq$2-jet requirement for
all three dilepton modes combined, compared to signal and background
predictions. The hatched bands reflect the uncertainties on the
background predictions.}\label{fig:njetsIdIsoNozMet_paper}
\end{center}
\end{figure}

Expected yields from simulated signal and background processes,
normalized to estimates from data where appropriate, are shown in
Fig.~\ref{fig:njetsIdIsoNozMet_paper} as a function of jet
multiplicity for events satisfying
the complete dilepton event
selection criteria except the $\geq$2-jet requirement; the
${\rm \ttbar}$ signal dominates the bins with at least two jets.

\begin{figure}[t!]
\begin{center}
  \includegraphics[width=0.45\textwidth]{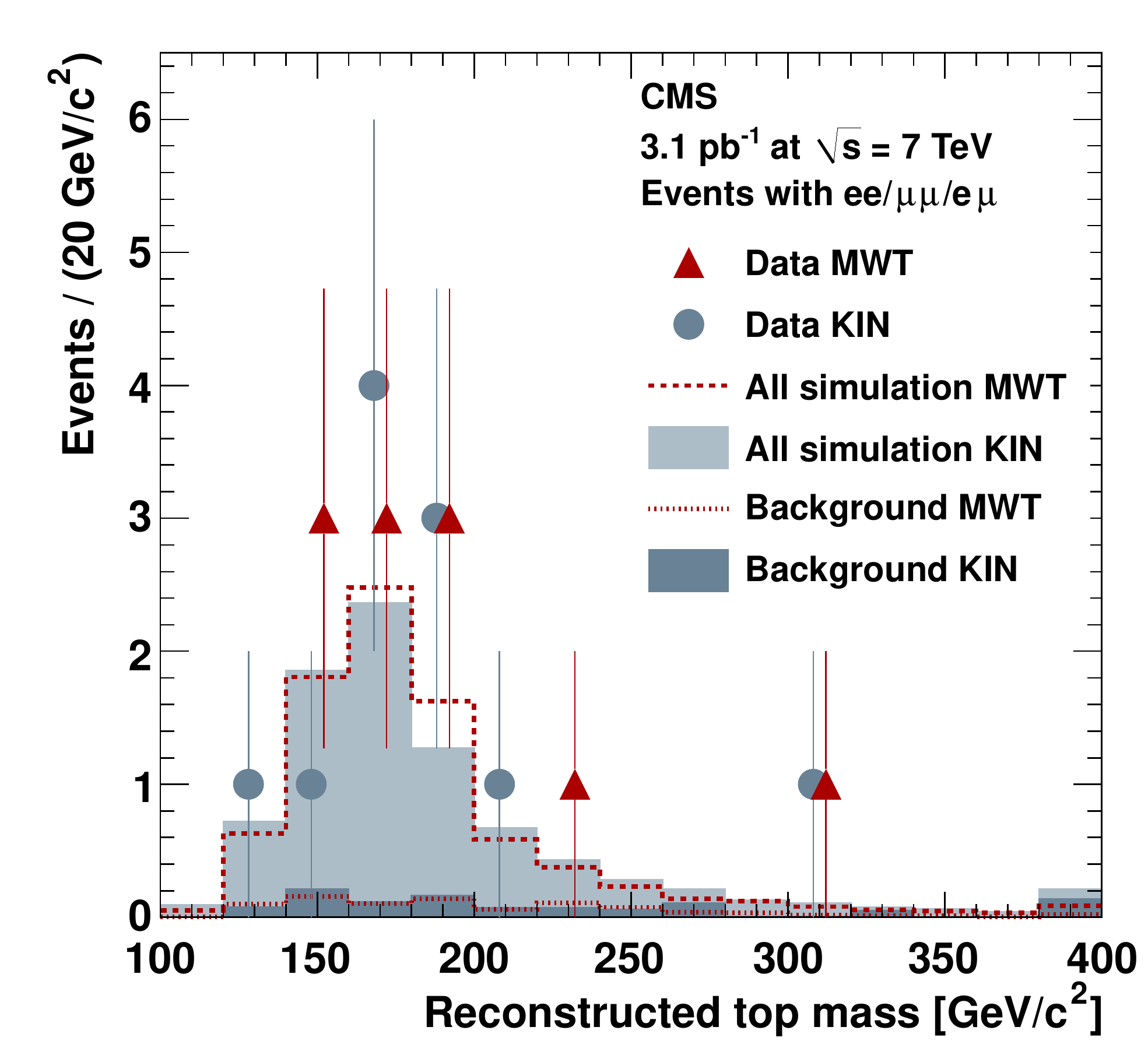}
\caption{\small 
Distribution of the top-quark mass using two different reconstruction
methods~\cite{topmassCDF,topmassD0}, compared with the expected yields
from simulated signal-plus-background and background-only
hypotheses. The points in each bin for the two methods are slightly offset in
reconstructed mass to allow coincident points to be visible. The last
bin contains the overflow.}\label{fig:topmass}
\end{center}
\end{figure}

Eleven dilepton events (3 ${\rm e^+e^-}$, 3 $\Pgmp\Pgmm$, 5 ${\rm
e^{\pm}}\Pgm^{\mp}$) are observed in the data after applying the
event selection criteria, with a total of $2.1{\rm \pm}1.0$ background
events expected. We attribute the excess of events above the
background expectation to top-quark pair production.

The top-quark mass reconstruction methods of
\cite{topmassCDF} (KIN, i.e., KINematic, method) and
\cite{topmassD0} (MWT, i.e., Matrix-element Weighting
Technique) are applied to the selected events.  In both methods,
numerical solutions to the kinematic equations appropriate for 
a ${\rm \ttbar}$ decay with two charged leptons in the final state
are found for each event.  The solutions are based 
on an ensemble of values of jet momenta and missing energy, generated
corresponding to their expected resolution around the measured values.
In the KIN method the underconstrained system is solved by
introducing an additional constraint on the longitudinal momentum of
the ${\rm \ttbar}$ system, whose probability distribution is expected
to have a negligible dependence on the top-quark mass and is therefore
assumed from simulation. The top-quark mass value corresponding to the
largest number of solutions is the reconstructed mass for each event.
In the MWT method the system is solved for a range of top-quark mass
values, and weights are assigned based on the likelihood of each
solution. The solution with the largest weight is used as the mass
estimator.  Figure~\ref{fig:topmass} shows that the
kinematics of the selected events are statistically compatible with
predictions based on a top-quark mass of 172.5~\GeVcc , demonstrating
the consistency of the selected sample with top-quark pair production.

\begin{figure}[t!]
\begin{center}
  \includegraphics[width=0.45\textwidth]{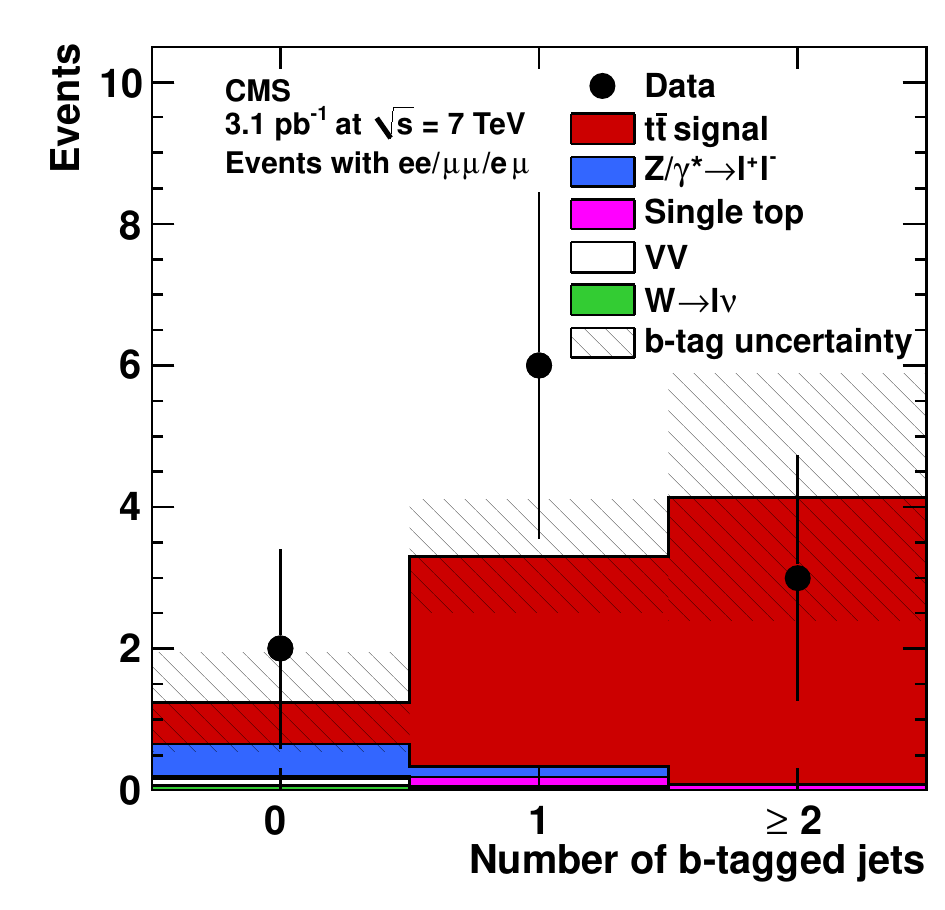}
\caption{\small Number of b-tagged jets in events passing
all dilepton selection criteria for all three dilepton modes combined,
compared to signal and background predictions. The hatched bands
reflect the expected uncertainties on the b-tag efficiency for signal
events.}\label{fig:btagsIdIsoNoz2jetsMet_paper}
\end{center}
\end{figure}

Further, beyond the complete event selection described above, the
property that the two jets expected in dilepton ${\rm t{\bar t}}$
events both originate from ${\rm b}$ quarks is exploited to 
further confirm the top-quark signal.  A ${\rm b}$-quark jet
identification algorithm that relies on the presence of charged
particle tracks displaced from the primary ${\rm pp}$ interaction
location, as expected from the decay products of long-lived ${\rm
b}$ hadrons~\cite{BTVPAS}, is used.  A jet is identified to be from a
${\rm b}$ quark if there are at least two tracks satisfying a minimum
impact parameter significance requirement.  The efficiency of this
algorithm for a ${\rm b}$-quark jet in dilepton ${\rm \ttbar}$ signal
events is about 80\% with a 10\% false positive rate, as estimated in
simulated QCD multijet events with no ${\rm b}$ quarks.  This
algorithm is applied to events passing all the selection criteria.
The multiplicity
of jets satisfying these ${\rm b}$-tagging criteria in events passing the full
dilepton event selection is shown in
Fig.~\ref{fig:btagsIdIsoNoz2jetsMet_paper}.  Although not used
directly in the cross section extraction, the b-tag multiplicity
provides additional support for the hypothesis that the selected data
are consistent with dilepton ${\rm \ttbar}$ production.

The top-quark pair production cross section is determined from the
ratio of the number of observed events in the data after background
subtraction with the product of the signal acceptance, selection
efficiency, the branching fractions, and the integrated luminosity.
From the simulated ${\rm \ttbar}$ sample, the acceptance times
efficiency is found to be ($23.0 \pm 1.4$)\% for events contributing
to the ${\rm e^+e^-}$, \Pgmp\Pgmm , and ${\rm e^{\pm}}$\Pgm $^{\mp}$
modes combined.  The total branching fraction for ${\rm \ttbar}$ to
the three modes of our final state is ($6.45 \pm 0.11$)\%~\cite{PDG}.
The systematic uncertainty on the acceptance times efficiency is
described below.



Various sources of systematic uncertainty related to the event
selection have been evaluated.
The systematic uncertainty assigned to the dilepton selection
efficiency is 4.4\%, obtained from a comparison of Z events in
data and simulation, together with half of the difference between the
efficiencies obtained in simulated Z and ${\rm t{\bar t}}$ events.
The effect of multiple pp interactions in a single beam crossing --- an
effect that is present in the data but not in these simulated samples
--- is included in this uncertainty. The systematic uncertainty due to
the reconstruction of jets and missing transverse energy is estimated
by varying the jet energy scale by ${\rm \pm}$5\%, simultaneously with
a ${\rm \pm}$5\% variation in the hadronic part of the missing
transverse energy, resulting in a value of 3.7\%.
Uncertainties on the simulation of the signal selection include the
amount of QCD radiation, hadron and tau decay modeling, and the W
leptonic branching fraction; these sources combined give a systematic
uncertainty of 2.8\%.  Other sources of systematic uncertainty
pertaining to the signal, including uncertainties in the parton
distribution functions inside the colliding protons and the effect of
additional minimum bias interactions in the signal selection, are
neglected because they were found to have a relatively small impact.
The overall systematic uncertainty on the total ${\rm t{\bar t}}$
cross section from the above sources is 6.4\%.

The background contributions from single-top, diboson, and Drell-Yan
${\rm Z}/\Pgg^\star \rightarrow \Pgt^+\Pgt^-$ processes shown in
Table~\ref{tab:yieldsAllCutsFinal2} are obtained from simulation and
found to be small compared to the total event yield. Each of these
backgrounds is assigned a 50\% systematic uncertainty. The
contributions from Drell-Yan ${\rm e^+e^-}$ and $\Pgmp\Pgmm$ processes
and events with non-W/Z isolated leptons are estimated from data with
absolute systematic uncertainties of 0.5 and 0.3 events, respectively.
The contribution to the systematic uncertainty on the cross section
from the uncertainties on the background estimates is 11\%.

The total systematic uncertainty on the measured cross section,
dominated by the uncertainty on the estimated background yield, is 24
pb.  An additional systematic effect of 21 pb, due to a 11\%
relative uncertainty on the integrated luminosity
measurement~\cite{lumipas}, is quoted separately.


Taking into account the data yield, the background estimation, the
branching fraction, the signal acceptance and efficiency, the
integrated luminosity, and all associated statistical and systematic
uncertainties, the top-quark pair production cross section is measured
to be

$$\sigma({\rm pp\rightarrow t{\bar t}}) = {\rm 194 \pm 72 (stat.) \pm 24 (syst.) \pm 21 (lumi.)~pb} .$$

\noindent
An alternative analysis, exploiting jets constructed only from silicon
tracker information~\cite{trkjetpas} and without missing transverse
energy requirements in the event selection, yields a similar 
cross section.  The quoted measurement can be compared with the
calculated NLO theoretical cross section of 157.5${\rm
^{+23.2}_{-24.4}}$ pb for a top-quark mass of 172.5~\GeVcc
~\cite{mcfm,mcfm:tt}.

In conclusion, the first measurement at the LHC of the cross section
for ${\rm t{\bar t}}$ production has been completed.  This
measurement, made with an integrated luminosity of ${\rm 3.1 \pm
0.3}$~\pbinv , is only the beginning of a rich top-quark physics
program to be conducted at the CMS experiment.


We wish to congratulate our colleagues in the CERN accelerator
departments for the excellent performance of the LHC machine. We thank
the technical and administrative staff at CERN and other CMS
institutes, and acknowledge support from: FMSR (Austria); FNRS and FWO
(Belgium); CNPq, CAPES, FAPERJ, and FAPESP (Brazil); MES (Bulgaria);
CERN; CAS, MoST, and NSFC (China); COLCIENCIAS (Colombia); MSES
(Croatia); RPF (Cyprus); Academy of Sciences and NICPB (Estonia);
Academy of Finland, ME, and HIP (Finland); CEA and CNRS/IN2P3
(France); BMBF, DFG, and HGF (Germany); GSRT (Greece); OTKA and NKTH
(Hungary); DAE and DST (India); IPM (Iran); SFI (Ireland); INFN
(Italy); NRF and WCU (Korea); LAS (Lithuania); CINVESTAV, CONACYT,
SEP, and UASLP-FAI (Mexico); PAEC (Pakistan); SCSR (Poland); FCT
(Portugal); JINR (Armenia, Belarus, Georgia, Ukraine, Uzbekistan); MST
and MAE (Russia); MSTD (Serbia); MICINN and CPAN (Spain); Swiss
Funding Agencies (Switzerland); NSC (Taipei); TUBITAK and TAEK
(Turkey); STFC (United Kingdom); DOE and NSF (USA).

\bibliography{auto_generated}




\cleardoublepage\appendix\section{The CMS Collaboration \label{app:collab}}\begin{sloppypar}\hyphenpenalty=5000\widowpenalty=500\clubpenalty=5000\textbf{Yerevan Physics Institute,  Yerevan,  Armenia}\\*[0pt]
V.~Khachatryan, A.M.~Sirunyan, A.~Tumasyan
\vskip\cmsinstskip
\textbf{Institut f\"{u}r Hochenergiephysik der OeAW,  Wien,  Austria}\\*[0pt]
W.~Adam, T.~Bergauer, M.~Dragicevic, J.~Er\"{o}, C.~Fabjan, M.~Friedl, R.~Fr\"{u}hwirth, V.M.~Ghete, J.~Hammer\cmsAuthorMark{1}, S.~H\"{a}nsel, C.~Hartl, M.~Hoch, N.~H\"{o}rmann, J.~Hrubec, M.~Jeitler, G.~Kasieczka, W.~Kiesenhofer, M.~Krammer, D.~Liko, I.~Mikulec, M.~Pernicka, H.~Rohringer, R.~Sch\"{o}fbeck, J.~Strauss, A.~Taurok, F.~Teischinger, W.~Waltenberger, G.~Walzel, E.~Widl, C.-E.~Wulz
% ...
\vskip\cmsinstskip
% ...
1:~~Also at CERN, European Organization for Nuclear Research, Geneva, Switzerland\\
% ...

\end{sloppypar}
\end{document}